\documentclass[prl,aps,amssymb,reprint,twocolumn,superscriptaddress,longbibliography,10pt]{revtex4-2}
\usepackage{project_macros-public}
\usepackage{project_macros-private}
\usepackage{makecell}

\crefname{appendix}{Appendix}{Appendices}
\crefname{equation}{Eq.}{Eqs.}
\crefname{figure}{Fig.}{Figs.}
\crefname{table}{Table}{Tables}
\crefname{section}{Section}{Sections}
\creflabelformat{appendix}{[#2#1#3]}

\makeatletter
\AtBeginDocument{\let\LS@rot\@undefined}
\makeatother

\makeatletter
\renewcommand\onecolumngrid{
\do@columngrid{one}{\@ne}%
\def\set@footnotewidth{\onecolumngrid}
\def\footnoterule{\kern-6pt\hrule width 1.5in\kern6pt}%
}

\global\long\def\bsl#1{\boldsymbol{{\bf #1}}}%

\global\long\def\ket#1{\left| #1\right\rangle }%

\global\long\def\bra#1{\left\langle #1 \right|}%

\global\long\def\up{\uparrow}%

\global\long\def\down{\downarrow}%

\allowdisplaybreaks

\newif\ifarxiv
\newif\ifprl

\arxivtrue 

\begin{document}
\title{\titlePaper}
\paperAuthors

\author{Miguel Gonçalves}
\affiliation{Princeton Center for Theoretical Science, Princeton University, Princeton NJ 08544, USA}

\author{Kun Yang}
\affiliation{Department of Physics and National High Magnetic Field Laboratory, Florida State University, Tallahassee, Florida 32306}

\author{Shi-Zeng Lin}
\affiliation{Theoretical Division, T-4 and CNLS, Los Alamos National Laboratory, Los Alamos, New Mexico 87545, USA}
\affiliation{Center for Integrated Nanotechnologies (CINT), Los Alamos National Laboratory, Los Alamos, New Mexico 87545, USA}

\let\oldaddcontentsline\addcontentsline

\ifarxiv
  \begin{abstract}
We show that chiral superconductivity can be stabilized by hole doping a Chern ferromagnet. Performing exact diagonalization and density-matrix-renormalization-group calculations on the repulsive Kane--Mele--Hubbard model at hole doping relative to filling $\nu=1$ electron per unit cell, we find that a Cooper pair formed by a magnon (spin-flip excitation) bound to two holes is stabilized at sufficiently strong interactions and sufficiently large Ising spin--orbit coupling (SOC). This Cooper pair exhibits both finite spin chirality---signaling a noncoplanar skyrmionic spin texture---and chiral $f$-wave symmetry. The pairing and spin chirality are set by the Chern number/polarization of the parent Chern ferromagnet. We further find that interactions between skyrmion Cooper pairs evolve from repulsive to attractive as the Ising SOC increases, revealing an intermediate-SOC region where chiral superconductivity can emerge from the condensation of hole-skyrmion Cooper pairs. Our findings provide a novel microscopic mechanism for chiral superconductivity and may be relevant for the recent observation of superconductivity in the MoTe$_2$ moir\'e superlattice.
\end{abstract}
\maketitle

\paragraph*{Introduction.---}
Unconventional superconductivity from repulsive electronic interactions has long been a central problem in condensed matter physics~\cite{Kohn1965SuperconductivityRepulsion}. The question has acquired renewed interest in moir\'e materials, where narrow isolated bands, nontrivial topology, and strong correlations place magnetism and superconductivity on comparable footing~\cite{Cao2020MoireFlatBandsReview}. Particularly striking is the recent experiment in twisted bilayer MoTe$_2$, which found superconductivity in the first moir\'e Chern band ~\cite{Xu2025MoTe2Experiment} in close proximity to re-entrant and fractional quantum anomalous Hall ferromagnets \cite{Park2023,Zeng2023,Cai2023,PhysRevX.13.031037,Ji2024,Redekop2024,Xu2025MoTe2Experiment,park2026observationhightemperaturedissipationlessfractional}. Together, these observations call for the exploration of microscopic pairing mechanisms that naturally intertwine topology and magnetism.

Several mechanisms for unconventional superconductivity have been proposed. In multiband systems, repulsion can generate effective attraction through interband polarization, correlated hopping, or excitonic processes~\cite{Crepel2021ExactTheoryRepulsiveSC,UnconventionalSCInterbandPolarization2021,Crepel2022TripletExcitonic,Crepel2023TopologicalSCDopedMagneticMoire,Jonas2024ElectronExcitonSC,guerci2025ferromagneticsuperconductivityexcitoniccooper}; repulsive spinless fermions with sublattice potentials can similarly yield effective attraction through virtual processes~\cite{RepulsiveSpinlessSublattice2023}. In moir\'e Chern bands, repulsion has also been argued to favor topological superconductivity in spin-polarized or spin-valley-polarized states, including recent proposals motivated by twisted WSe$_2$ and twisted MoTe$_2$~\cite{Guerci2024TwistedWSe2,ChiralSCSpinPolarizedChernMoTe2,Chen2025FiniteMomentumMoTe2,DongLee2025FwaveSpinPolarized,TopologicalSCEmergentVortex2026,zm39-dstj,fcdc-9lm3,Wang_Zaletel_2025,Dong_Lee_2026,gt8h-czf3}. Under short-range attraction, chiral $p_x+ip_y$ superconductivity has been demonstrated in rhombohedral graphene~\cite{BerryTrashcanRhombohedralGraphene2025} and in a two-dimensional electron gas of spin-polarized electrons~\cite{AttentionSolveChiralSC2025}.
In parallel, a significant body of work has highlighted that the elementary charge carriers of doped ferromagnets need not be bare electrons or holes. In frustrated triangular-lattice systems, a doped carrier can bind to a spin flip and form a spin polaron~\cite{Jiang2018PairingStrongRepulsionTriangular,Wang2023ItinerantSpinPolaron,Seifert2024SpinPolaronsFerromagnetism}, which has been already experimentally observed~\cite{Tao2024ObservationSpinPolarons}. In $SU(2)$ Chern ferromagnets (with application to twisted bilayer graphene), charged `baby' skyrmion excitations were proposed as arising from an effective attractive interaction between a magnon and an added electron charge \cite{Khalaf2022BabySkyrmionsChernFM}, similar to the mechanism for skyrmion formation in quantum Hall ferromagnets \cite{PhysRevB.47.16419,PhysRevLett.76.2153, GirvinYang2019}. Spin polarons have also been found in one-dimensional flat band ferromagnets with trivial band topology~\cite{SpinPolaronsFlatBandFerromagnets2025}. Spin-polaron-mediated superconductivity has also been proposed \cite{PhysRevLett.60.944,Jiang2018PairingStrongRepulsionTriangular,MagnonicSuperconductivity2024,Zhang_Batista_Zhang_2025,Zhu_Zhang_Zhang_Zhang_Batista_2026}, and was studied recently in doped Chern antiferromagnets~\cite{Wang2024IntertwinedFlatBands,Wang2025SpinPolaronCAF} and TBG \cite{Khalaf2021ChargedSkyrmionsMATBG,skyrmion_SC_zaletel,SkyrmionsTBGPairingCrystallization2022}.

\begin{figure}[h]
  \centering
  \includegraphics[width=\columnwidth]{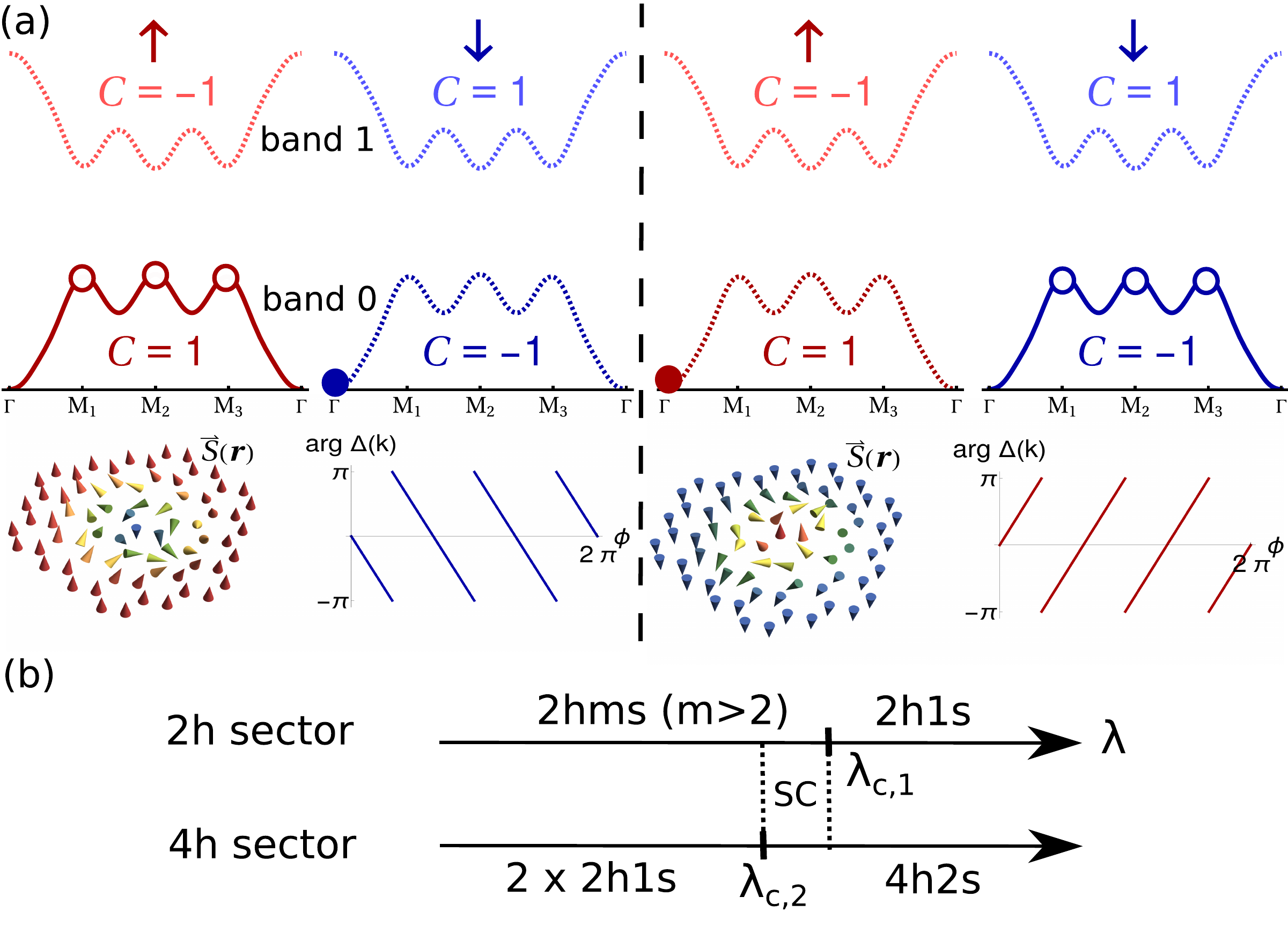}
  \caption{Summary of main results. (a) Illustration of the \textit{2h1s} skyrmion-bipolaron ground-state. {Filled (empty) bands are sketched using full (dashed) lines along a path in the Brillouin zone containing the $\Gamma$ and $\textbf{M}$ points. The picture illustrated here with absolute maxima/minima at the $\textbf{M}$ and $\Gamma$ points is valid for $0.2\lesssim |\lambda| \lesssim 0.575$}. Empty (filled) circles represent doped holes (electrons). $\Delta({\bf k})$ corresponds to the pair wave function and the plots illustrate its phase winding for a closed loop around ${\bf k} = 0$.  The sign of both the spin and pairing chiralities is set by the Chern number/polarization of the parent Chern ferromagnet. (b) Sketch of the phase diagram as a function of the Ising SOC $\lambda$, for a fixed interaction strength. }
  \label{fig:1}
\end{figure}

This raises two important open questions. In a doped \emph{Chern ferromagnet}, can the relevant Cooper pairs be formed not from fully spin-polarized carriers, but from composite \textit{spinful} carriers generated by binding charge to spin-flip excitations? And if such pairing occurs, does the resulting superconducting state remain topological by virtue of its origin in a Chern band? This is well motivated by known results: (i) in the absence of topology, spin bipolarons have been proposed as stable Cooper pairs~\cite{Jiang2018PairingStrongRepulsionTriangular,MagnonicSuperconductivity2024,zhu2026magnonmediatedsuperconductivityinfiniteutriangular}; (ii) band mixing has been observed to render spinful excitations energetically competitive in multiband topological systems, and in particular, in models of twisted MoTe$_2$~\cite{MFCI0,kwan2024abelianfractionaltopologicalinsulators,goncalves2025spinlessspinfulchargeexcitations}. 
In fact, this question is particularly relevant in the context of Ising (spin-$U(1)$) Chern ferromagnets like MoTe$_2$, where bands with opposite spin polarization have opposite Chern numbers. In this case, even the formation of spin-polarons has remained largely unexplored. It was shown that skyrmion excitations can be stabilized upon doping at the mean-field level \cite{Goncalves2025DopingInducedQAH}, while a quantum geometric dipole mechanism responsible for the binding of opposite-spin electron and hole (magnon) excitations  was proposed in Ref.~\cite{chen2025quantumgeometricdipoletopologicalboost}.

Here we show that composite Cooper pairs formed from equal-spin holes bound to a spin-flip excitation can be stabilized by hole doping an Ising Chern ferromagnet. Using exact diagonalization (ED) and DMRG at hole doping relative to the $\nu=1$ Chern ferromagnet, we find that sufficiently strong interaction and Ising spin--orbit coupling stabilize a bound state of two holes and one magnon. This composite has a charge-$2e$ and spin $S_z=\mp2$, with $-(+)$ respectively relative to the spin-$\up$($\down$) Chern ferromagnet \footnote{For the spin-$\up$ Chern ferromagnet, $S_z=-2$ arises from $\Delta S_z=-1$ due to the two doped spin-$\up$ holes plus $\Delta S_z=-1$ from the spin-flip.}. This Cooper pair exhibits chiral $f$-wave symmetry and carries finite spin chirality. 
Both the pairing symmetry and spin chirality are set by the polarization/Chern number of the parent Chern ferromagnet, as illustrated in Fig.~\ref{fig:1}. We further show that the interaction between these hole-magnon pairs evolves from repulsive to attractive as the Ising spin--orbit coupling increases, identifying an intermediate SOC regime in which their condensation can realize chiral superconductivity. 
Crucially, while adding a finite Zeeman field is essential for stabilizing spin polarons in the mechanism previously described in Refs.~\cite{Jiang2018PairingStrongRepulsionTriangular,MagnonicSuperconductivity2024,zhu2026magnonmediatedsuperconductivityinfiniteutriangular}, in the present case the spin polarons form spontaneously at zero field.
Our results provide a concrete microscopic route from a doped Chern ferromagnet to chiral superconductivity, and show that multiband mixing can stabilize Cooper pairs as the lowest charge-$2e$ excitation by binding spin flips.

\paragraph*{Model and method.---}
We consider the repulsive Kane--Mele--Hubbard model given by
\begin{equation}
  \begin{aligned}
  H &= -t \sum_{\langle ij \rangle,\sigma} \left( c^\dagger_{i\sigma} c_{j\sigma} + \mathrm{H.c.} \right) 
  + \mathrm{i}\lambda \sum_{\langle\!\langle ij \rangle\!\rangle , \sigma} \sigma \nu_{ij} \left( c^\dagger_{i\sigma} c_{j\sigma} + \mathrm{H.c.}\right) \\
  &\quad + U \sum_i n_{i\uparrow} n_{i\downarrow},
  \end{aligned}
\end{equation}
where $\langle ij \rangle$ and $\langle\!\langle ij \rangle\!\rangle$ denote nearest and next-nearest neighbors on the honeycomb lattice, $c^\dagger_{i\sigma}$ creates an electron with spin $\sigma=\uparrow(+1),\downarrow(-1)$ at site $i$, $n_{i\sigma} = c^\dagger_{i\sigma} c_{i\sigma}$, $n_i = n_{i\uparrow} + n_{i\downarrow}$, and $\nu_{ij} = \pm 1$ is the sign of the next-nearest-neighbor hop (clockwise or counterclockwise around the hexagon). The second term is the Ising spin--orbit coupling, that also opens a topological gap for each spin species.
We study this problem using ED and DMRG. Because it requires two bands per spin calculation which severely restricts the available system sizes, we will use the "band
maximum" technique introduced in Ref.~\cite{yu2024moirefractionalcherninsulators} for the ED calculations, where the number of particles allowed in the highest-energy band per spin (band 1) is restricted to a number $N_1$ and the result is converged for increasing $N_1$, while the number of particles in the lowest band (band 0) is unrestricted.
DMRG calculations were performed using a quasi-1D geometry, with 6 sites in the transverse direction along which periodic boundary conditions are imposed, and a variable number of sites in the longitudinal direction up to 24, along which open boundary conditions were employed. 
In all our calculations, we conserve both the total number of particles and total $S_z$.

\begin{figure}[h!]
  \centering
  \includegraphics[width=\columnwidth]{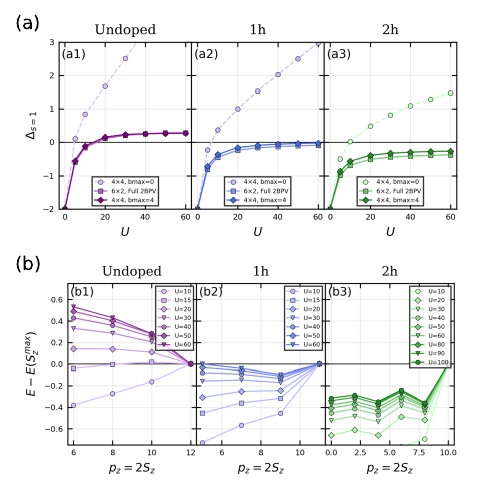}
  \caption{Exact diagonalization results for $\lambda=0.225$ on a 12-unit-cell lattice with $N_1=4$. (a) Spin-flip gap with respect to the fully polarized state as function of interaction strength $U$, for different doping levels indicated in the figure. (b) Ground-state energy vs. $p_z=2 S_z$ for different $U$ and different doping levels. }
  \label{fig:2}
\end{figure}

\paragraph*{Skyrmion-polaron ground-state.---}
We start by showing that there is a $U$-induced ferromagnetic transition. The critical interaction strength can be estimated from the spin gap defined as $\Delta_s=E(N,N/2-s)-E(N,N/2)$, where $E(N,S_z)$ is the ground-state energy for $N$ particles (equal to the number of unit cells at this filling $\nu=1$) and spin sector $S_z$. A lower-bound for the critical interaction strength can be estimated from $\Delta_{s=1}$ shown in Fig.~\ref{fig:2}(a1), which can be well converged in system size. We further show in Fig.~\ref{fig:2}(b1) that for large enough interaction strength, the spin-polarized Chern ferromagnet is the ground-state compared to all the accessible spin sectors. The critical interaction strength for the ferromagnetic transition $U_c$ decreases with $\lambda$ for the range of parameters studied ($\lambda\geq0.1$), approximately stabilizing at $\lambda \gtrsim 0.2$ (see Supplemental Materials (SM) \cite{SM}). For $U>U_c$, the magnon excitations on top of the Chern ferromagnet are gapped due to the broken $SU(2)$ symmetry. However, quite remarkably, when a hole is doped with respect to the Chern ferromagnet, it becomes energetically favorable for it to bind with a magnon and form a \textit{1h1s} spin polaron (see SM for explicit binding energy calculations  \cite{SM}). This is explicitly shown in Fig.~\ref{fig:2}(a2), where we see that the (spin $|S_z|=3/2$) \textit{1h1s} polaron has smaller energy than the \textit{1h} state even for very large interaction strength. Such a result contrasts significantly with a one band per spin calculation, where the Hamiltonian is projected into the lowest-energy Chern bands per spin only. These results are also shown in Fig.~\ref{fig:2}(a2), showing that the \textit{1h1s} excitation has an energy cost proportional to $U$ -- i.e. the simpler \textit{1h} state, which can avoid $U$, becomes the ground-state for a large enough $U$. The crucial point is that because we are not at half-filling, spinful excitations are not suppressed by Pauli blocking. In particular, the projection of the Hamiltonian into the subspace with no double occupancy does not vanish, in contrast to the half-filled case, implying that we can have non-trivial kinetic magnetism even when $U\rightarrow \infty$. As we will see later, the \textit{1h1s} polaron has a finite spin chirality whose sign is determined by the Chern number/polarization of the parent Chern ferromagnet. This implies that it has a non-coplanar skyrmion-like spin texture akin to the `baby skyrmions' introduced in Ref.~\cite{Khalaf2021ChargedSkyrmionsMATBG}. We will then label this bound-state as a \textit{skyrmion-polaron} composite, to contrast with the more conventional polaron, that has a vanishing spin chirality. 
In Fig.~\ref{fig:2}(b2) we can show that the \textit{1h1s} polaron is the stable ground-state for the range of accessible $S_z$ (convergence in $N_1$ for smaller values of $S_z$ becomes unfeasible due to the large required Hilbert-space dimension). In particular, it is stabilized for larger $U$ [Fig.~\ref{fig:2}(b2)] and Ising SOC $\lambda$ \cite{SM}. Although the results in Fig.~\ref{fig:2}(b) are shown for a fixed value of $N_1$, this value is sufficient to guarantee the convergence of the results \cite{SM}.

\paragraph*{Skyrmion-bipolaron ground-state.---}
We now explore the ground-state for two doped holes with respect to the Chern ferromagnet. In Fig.~\ref{fig:2}(a3) we show that the bipolaron formed by two holes and a magnon has lower energy than the simple two-hole state even for very large $U$. This necessarily implies that the extra hole that is added binds to the \textit{1h1s} polaron, forming a \textit{2h1s} spin-bipolaron (see SM for explicit binding energy calculations \cite{SM}). 
We also find that for large enough $U$ [Fig.~\ref{fig:2}(b3)]  and $\lambda$ \cite{SM}, the spin bipolaron can be stabilized as the absolute ground-state. 
{This can be partly understood from a mechanism similar to that proposed in Ref. \cite{MagnonicSuperconductivity2024} for trivial bands. In particular, for $|\lambda|\gtrsim 0.2$, the lowest-energy band develops maxima at the three inequivalent $\mathbf{M}$ points and a minimum at the $\Gamma$ point. As a result, it becomes energetically favorable to bind three holes at the $\mathbf{M}$ points in one spin sector with an electron at $\Gamma$ in the opposite spin sector, as illustrated in Fig.$,$\ref{fig:1}(a). We emphasize, however, that virtual processes involving the higher-energy bands are essential to stabilize this skyrmion bipolaron as the ground state, as previously shown in Fig.$,$\ref{fig:2}(a3).}
Because the \textit{2h1s} spin bipolaron also has a finite spin chirality, we name it a \textit{skyrmion-bipolaron} composite.

\begin{figure}[b]
  \centering
  \includegraphics[width=\columnwidth]{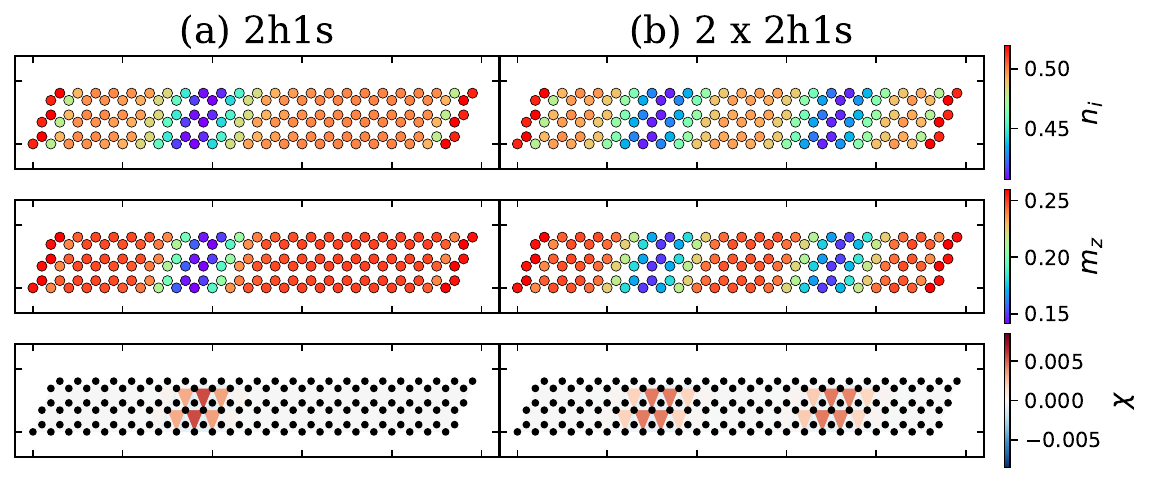}
  \caption{DMRG results for $\lambda=0.225$ and $U=100$ in the \textit{2h1s} and \textit{4h2s} sectors. Top row: charge density $n_i$. Middle: magnetization $m_z$. Bottom: spin chirality $\chi_{ijl}$. }
  \label{fig:3}
\end{figure}

\paragraph*{Interaction between Skyrmion-bipolarons.---}
We now focus on the interaction between two skyrmion-bipolarons. Because this problem is unfeasible in ED for system sizes with $N_{UC}\geq 12$ unit cells, due to Hilbert-space size limitations, we turn to DMRG. We start by showing the results for the skyrmion-bipolaron composite in Fig.~\ref{fig:3}(a). Because translation symmetry is broken in the longitudinal direction due to open boundaries, we can see a localized depletion in the charge density in Fig.~\ref{fig:3}(a1), corresponding to the region in space where the hole is doped with respect to the ferromagnetic background. Exactly in the same region, we see a depletion in $m_z$ in Fig.~\ref{fig:3}(a2), as expected for a bound-state between two holes and a magnon. Finally, in Fig.~\ref{fig:3}(a3) we plot the scalar spin chirality defined as $\chi_{ijl} = \langle S_i \cdot (S_j \times S_l) \rangle$, where the site indices $(i,j,l)$ are defined in the different triangular plaquettes of the honeycomb lattice. It is clear that a finite spin chirality develops precisely at the position of the skyrmion-bipolaron composite, while vanishing everywhere else. It is important to note that the sign of the total spin chirality is determined by the Chern number/polarization of the parent Chern ferromagnet, see SM \cite{SM}. In the SM, we also show the DMRG results for the skyrmion-polaron composite, where a clear binding between the hole and the magnon can also be observed. \cite{SM}

To determine the interaction between two skyrmion-bipolarons, we obtain  DMRG results in the sector of 4 holes and 2 spin-flips with respect to the Chern ferromagnet. We find that there is a critical value of $\lambda$ below which the interaction between two skyrmion-bipolarons is repulsive, and above which it is attractive. This can be seen directly in the DMRG results that either converge to a ground-state with two separated skyrmion-bipolarons, or a ground-state with a single larger \textit{4h2s} skyrmion-bipolaron bound-state. In Fig.~\ref{fig:3}(b) we show the results for the former case, for the same choice of $\lambda = 0.225$ used in Fig.~\ref{fig:2}, where \textit{2h1s} is the ground state. The interaction becomes attractive above $\lambda_{c,2}=0.23$ \cite{SM}. 
The DMRG and ED results combined therefore indicate that there can exist a small intermediate-SOC regime where the \textit{2h1s} skyrmion-bipolaron is the ground-state and the skyrmion-bipolarons repel, as illustrated in Fig.~\ref{fig:1}(b). We explicitly checked that for $U=100$, we have $\lambda_{c,1}=0.222$, and therefore for  $\lambda_{c,1} < \lambda < \lambda_{c,2}$, a dilute concentration of the \textit{2h1s} bosons can  condense to form a  superconductor. We note that while the parameter range for the intermediate-SOC regime is small, it depends on model details. The aim of this work is not to make an exhaustive parameter model scan of when this regime can be stabilized, but instead to establish that it can be energetically favored even in the paradigmatic Kane-Mele-Hubbard model. Equally important, we also aim to understand the mechanism for superconductivity in this regime, which will be the subject of the remainder of the manuscript. 

\begin{figure}[h!]
  \centering
  \includegraphics[width=\columnwidth]{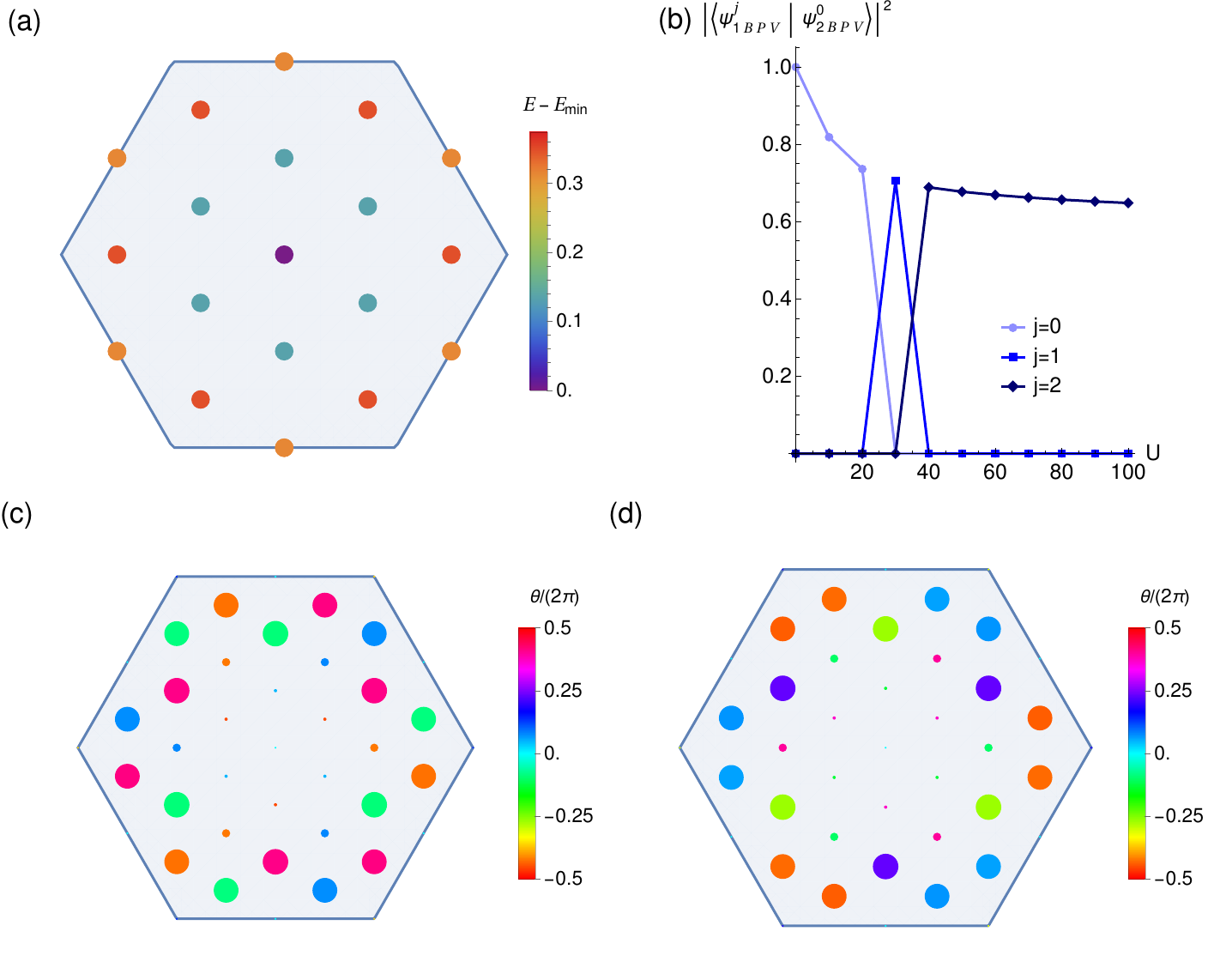}
  \caption{Energy dispersion and pairing symmetry of the \textit{2h1s} Cooper pair. (a) Dispersion for a 16-unit-cell lattice and  $N_1=4$. (b) Overlap of the full two-band ground-state and the three one-band projected lowest-energy states as a function of $U$. (c) $\Delta(\bf{k})$ for the $\ket{\Uparrow}$ Chern ferromagnet parent state, for a 36-unit-cell lattice. The size of the points are proportional to $|\Delta(\bf{k})|$, while the phase $\theta(\bf{k})$ is given by the color map. (d)  $\Delta(\bf{k})$ for the $\ket{\Downarrow}$ Chern ferromagnet parent state.  }
  \label{fig:4}
\end{figure}

\paragraph*{Pairing symmetry.---}
Having established that the \textit{2h1s} composite can be an energetically stable Cooper pair, we now focus on understanding the pairing symmetry of the superconducting state arising from the condensation of these Cooper pairs. 
We first plot the dispersion of the  \textit{2h1s} cooper pair in the Brillouin zone in Fig.~\ref{fig:4}(a) [see SM \cite{SM} for additional plots of the magnon and spin-bipolaron dispersions]. We can see that there is a substantial dispersion away from the minimum at the $\Gamma$ point, even when $U\rightarrow \infty$. This supports the condensation of the Cooper pair in the dilute density limit. 

We now turn to establish the pairing symmetry of the superconducting state. This can be partly answered by inspecting the $C_6$ eigenvalue of the \textit{2h1s} excitation. The ground-state in the \textit{2h1s} sector can be written as $\ket{\Uparrow + \textrm{2h1s}} = \mathcal{O} \ket{\Uparrow}$, where $\mathcal{O}$ is the operator that creates the \textit{2h1s} composite on top of the Chern ferromagnet $ \ket{\Uparrow}$, explicitly given by $ \ket{\Uparrow}=\prod_{\bf k} \gamma^{\dagger}_{0,\up,{\bf k}} \ket{0}$ in the one-band limit. The $C_6$ eigenvalue of this excitation is then given by $\lambda_{\mathcal{O}}=\lambda_{\Uparrow + \textrm{2h1s}}/\lambda_{\Uparrow}$. We consistently find $\lambda_{\mathcal{O}}=-1$ for the 12- and 16-unit-cell lattices, indicating pairing $f$-wave symmetry. 
However, just by the $C_6$ eigenvalue we cannot establish whether the $f$-wave pairing is chiral or not. This question is challenging to answer for the small system sizes we have access in the two-band calculation. We therefore turn to ask if the \textit{2h1s} $f$-wave state is adiabatically connected to the $f$-wave state obtained in the one-band limit, where we project only to the lowest-energy band per spin. We address this question by plotting in Fig.$\,$\ref{fig:4}(b) the overlap of the \textit{2h1s} $f$-wave ground-state in the two-band calculation with the lowest-energy states in the one-band calculation, as a function of interaction strength. We can see that as $U$ increases, the $f$-wave state becomes an excited state in the one-band calculation. However, the states that are lower in energy have different $C_6$ eigenvalues, and therefore are crossed by the $f$-wave state without mixing. In fact, if we track the one-band $f$-wave state as function of $U$, we see that the overlap with the 2-band $f$-wave state is always above 0.65 and does not show any discontinuity -- indicating that the one-band and two-band $f$-wave states are adiabatically connected. Since the pairing symmetry should not change as long as the two states are adiabatically connected, we conclude about the symmetry of the two-band $f$-wave Cooper pair by inspecting the symmetry of the one-band pair. In the one-band limit, we can formulate a simple definition for the gap order parameter, through 
\begin{equation}
 \Delta({\bf k},\uparrow) = \bra{\Uparrow + \textrm{2h1s}} S_0 \gamma_{0,\up,{\bf k}} \gamma_{0,\up,-{\bf k}} \ket{\Uparrow}
 \label{eq:gap_Deltak},
\end{equation}
where $\gamma_{0,\up,{\bf k}}$ creates a hole in band 0 with spin up at momentum ${\bf k}$ and $ S_0$ is a zero-momentum local spin flip operator defined as $S_0 = \sum_{\bf{k},\alpha} \langle u^{\alpha}_{0,\down,{\bf k}}  | u^{\alpha}_{0,\up,{\bf k}}  \rangle \gamma^{\dagger}_{0,\down,{\bf k}} \gamma_{0,\up,{\bf k}}$, where $u^{\alpha}_{0,\eta,{\bf k}}$ is the band-0 single-particle eigenstate amplitude on sublatice $\alpha$ with spin $\eta$ and momentum ${\bf k}$. In Fig.~\ref{fig:4}(c) we plot $\Delta({\bf k})=|\Delta({\bf k})|e^{i\theta({\bf k})}$, where we can see the $f$-wave node structure clearly and that the phase $\theta({\bf k})$ winds around ${\bf k}=0$. In particular, if we consider a loop around the $\Gamma$ point, we see that the phase $\theta({\bf k})$ winds three times, showing $f_x-if_y$ symmetry. If we, on the other hand, consider the \textit{2h1s} state formed by doping the spin-down Chern ferromagnet -- i.e. replacing $\up \rightarrow \down$ and $\Uparrow \rightarrow \Downarrow$ in Eq.$\,$\ref{eq:gap_Deltak}, we see that the phase has the opposite winding, indicating $f_x+if_y$ symmetry, as shown in Fig.~\ref{fig:4}(d).

\paragraph*{Discussion.---} 
We have established the stabilization of chiral superconductivity by doping holes in a Chern ferromagnet. 
The main new ingredient behind our mechanism is the skyrmion-bipolaron Cooper pair. Previous works have established the possibility of magnonic superconductivity, which can be stabilized for a large enough Zeeman field. The corresponding Cooper pair is however obtained by doping a topologically trivial ferromagnet, resulting in a trivial spin texture. Our mechanism allows for the stabilization of the skyrmion-bipolaron Cooper pair spontaneously in the zero magnetic field. 
Our mechanism also differs from the recent proposal of chiral superconductivity obtained by doping holes in a Chern band in Refs.$\,$\cite{Guerci2024TwistedWSe2,ChiralSCSpinPolarizedChernMoTe2,zm39-dstj}, where no spinful excitations are present. Our work sheds a new light on the superconducting phase observed in twisted MoTe$_2$ in Ref.~\cite{Xu2025MoTe2Experiment}. In MoTe$_2$, band mixing has been recently shown to make spin-flip and spin-polarized excitations energetically competitive, and therefore it would be natural to have regimes where the skyrmion-bipolaron Cooper pair is the ground state.
We finally comment on the connection of our results with earlier proposals of skyrmion superconductivity in Refs.$\,$\cite{Khalaf2021ChargedSkyrmionsMATBG,SkyrmionsTBGPairingCrystallization2022,skyrmion_SC_zaletel}. In earlier proposals, superconductivity arises due to the binding between skyrmion--anti-skyrmion pairs originating from doping each charge in a band with opposite Chern number. Additionally, the
parent state is a $C=0$ Kramers intervalley-coherent state, that breaks valley-$U(1)$ symmetry spontaneously.
Our parent state, on the other hand, is a Chern ferromagnet. The non-coplanar spin texture is only induced upon doping.  

While in our work we have established the possibility of having a Cooper pair of two-holes bound to a single spin flip, our results also suggest that skyrmion-bipolarons with higher spin can also be stabilized in different regimes. In particular, while the \textit{2h1s} skyrmion-bipolaron is the ground-state for a small range of parameters, it only becomes unfavored with respect to skyrmion-bipolarons with larger spin-flips, rather than with respect to the \textit{2h} state. Although higher-spin bipolarons are much more challenging to characterize using exact calculations due to the much larger associated Hilbert-space sizes, we expect that they will be described by the same fundamental ingredients we unveiled here.

\paragraph*{Acknowledgments.---}
Exact diagonalization computations were performed using the DiagHam library. DMRG computations were performed using the ITensor library.
We thank Pok Man Tam, Jonah Herzog-Arbeitman, Felipe Mendez, Danielle Guerci and Khachatur Nazaryan  for fruitful discussions. MG is supported by a postdoctoral research fellowship
at the Princeton Center for Theoretical Science. SZL is partially supported by the U.S. Department of Energy (DOE) National Nuclear Security Administration (NNSA) under Contract No. 89233218CNA000001 through the Laboratory Directed Research and Development (LDRD) Program and was performed, in part, at the Center for Integrated Nanotechnologies, an Office of Science User Facility operated for the DOE Office of Science, under user Proposals No. 2018BU0010 and No. 2018BU0083. The work of KY is supported by the National Science Foundation Grant No. DMR-2315954, and performed at the National High Magnetic Field Laboratory, which is supported by National Science Foundation Cooperative Agreement No. DMR-2128556, and the State of Florida.

\newpage

\begin{center}
	\textbf{\large End Matter }\\[.2cm]
\end{center}

\paragraph*{Details on DMRG calculations.---} In all the DMRG calculations calculations we use a pinning field on the boundary sites given by $\delta H = \mp \sum_{i} S^z_i$, respectively for the $\ket{\Uparrow}$ and $\ket{\Downarrow}$ parent Chern ferromagnet states, which is necessary to prevent pinning the spin polarons at the boundary.
We use the strategy of running different DMRG runs for a fixed set of parameters. We only stop the calculation when (i) the relative error of the average entanglement entropy, $\overline{S_j}$, the average density, $\overline{|\langle n_j \rangle|}$ and the average energy is below a threshold of $\epsilon_{\textrm{obs}}$, where $\overline{\cdot}$ denotes an average over all sites; or (ii) the number of sweeps reaches above 250. To guarantee the convergence of our results, we run 3 different calculations with different tolerances given by $\epsilon_{\textrm{obs}} = 10^{-4}, 7.5 \times 10^{-5}, 5 \times 10^{-5}$, respectively with DMRG truncation errors of $\epsilon_{\textrm{trunc}} = 7.5 \times 10^{-6}, 5\times10^{-6}, 2.5\times10^{-6}$. Additionally, to avoid convergence into local minima, we run at least 10 different DMRG runs for each tolerance/truncation error, starting from different random initial states.

\fi

\renewcommand{\addcontentsline}[1]{}
\nocite{REVTEX42Control,apsrev42Control}
\bibliographystyle{apsrev4-2}
\bibliography{revtex-control,refs}
\let\addcontentsline\oldaddcontentsline

\renewcommand{\thetable}{S\arabic{table}}
\renewcommand{\thefigure}{S\arabic{figure}}
\renewcommand{\theequation}{S\arabic{section}.\arabic{equation}}
\onecolumngrid
\pagebreak
\thispagestyle{empty}

\cleardoublepage

\begin{center}
	\textbf{\large Supplementary Information for ``\titlePaper{}"}\\[.2cm]
\end{center}

\appendix
\setcounter{secnumdepth}{3} 
\renewcommand{\thesection}{\Roman{section}}
\tableofcontents
\let\oldaddcontentsline\addcontentsline
\newpage
\section{Details on the ED calculations}

\label{sec:App_details_ED}

\begin{figure}[h!]
  \centering
  \includegraphics[width=0.75\columnwidth]{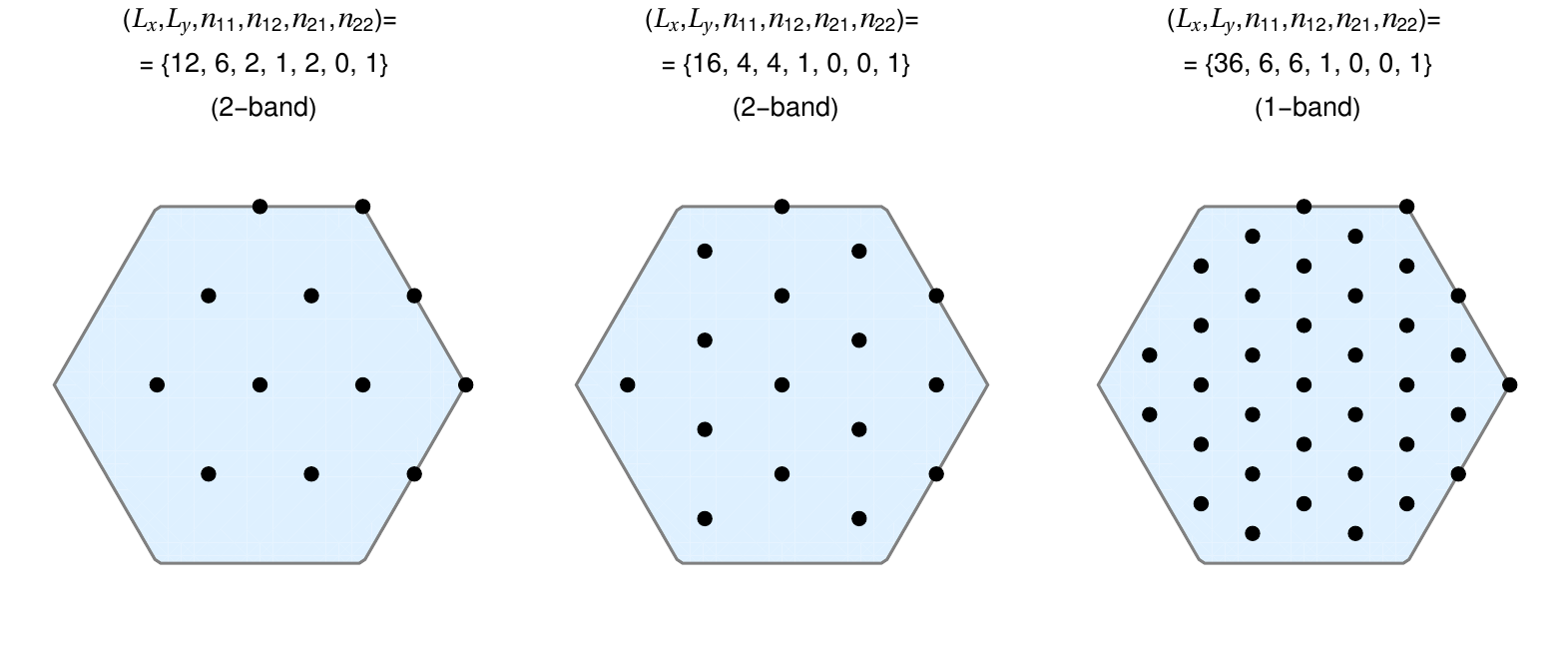}
  \caption{Momentum meshes used for the ED calculations. }
  \label{fig:momentum_meshes}
\end{figure}

In this Appendix section we provide additional details on the ED calculations presented in the main text.
We consider momentum meshes defined by the two reciprocal lattice vectors $\bsl f_{1}$ and $\bsl f_{2}$ given by
\begin{equation}
\begin{array}{cc}
\bsl f_{1} & =n_{11}\bsl b_{1}+n_{12}\bsl b_{2}\\
\bsl f_{2} & =n_{21}\bsl b_{1}+n_{22}\bsl b_{2}\,.
\end{array}
\end{equation}
where $\bsl b_{1}=(-\sqrt{3},1)/2$ and $\bsl b_{2}=(\sqrt{3},1)/2$ are the reciprocal
lattice vectors and $n_{11},n_{12},n_{21},n_{22}$ are integers satisfying
$|n_{11}n_{22}-n_{12}n_{21}|=1$. We take momentum meshes of $N_{x},N_{y}$
momenta along the directions $\bsl f_{1}$ and $\bsl f_{2}$ respectively,
that is,

\begin{equation}
\bsl k=\frac{k_{x}}{L_{x}}\bsl f_{1}+\frac{k_{y}}{L_{y}}\bsl f_{2}\,,
\end{equation}
where $k_{x}=0,\cdots,L_{x}-1$ and $k_{y}=0,\cdots,L_{y}-1$, corresponding
to a system size of $N_{UC}=L_{x}L_{y}$ unit cells. The momentum meshes used for the ED calculations are explicitly shown in Fig.$\,$\ref{fig:momentum_meshes}. Note that these meshes are all $C_6$ symmetric.

\section{"Band maximum" technique and convergence of results}
\begin{figure}[h!]
  \centering
  \includegraphics[width=0.75\columnwidth]{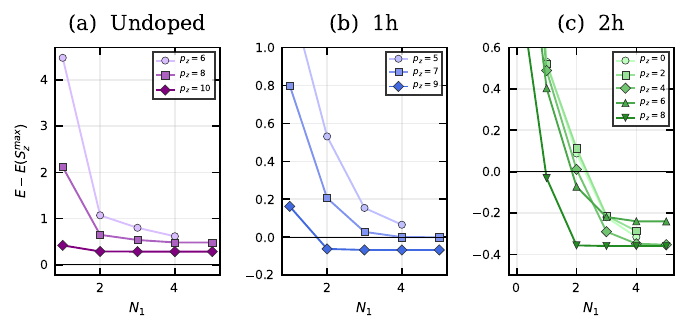}
  \caption{Energy difference with respect to the fully polarized state for different $p_z=2S_z$ as a function of number of particles in band $1$, $N_1$, for different doping levels with respect to $\nu=1$: (a) Undoped; (b) one doped hole; (c) two doped holes. All results are shown for the 12-unit-cell lattice, $U=100$ and $\lambda=0.225$. }
  \label{fig:App_convergence_bmax}
\end{figure}

Some results presented in the main-text were obtained using the "band
maximum" technique introduced in Ref.~\cite{yu2024moirefractionalcherninsulators}, which amounts to restricting the occupation of the highest-energy band per spin (band~1) to at most $N_1$ particles.
The advantage of this technique is that it allows reaching larger system sizes than the ones accessible in the full two-band calculation by truncating the Hilbert space size. In our current problem, we are considering cases where the interaction strength $U$ can be much larger than the topological gap between bands 0 and 1. Therefore, the convergence of the "band maximum" technique in this case needs to be carefully checked. In 
Fig.~\ref{fig:App_convergence_bmax} we explicitly plot how some of the energies shown in the main text converge with increasing $N_1$. Where we could reach sufficiently large Hilbert-space sizes to compare $N_1=4$ and $N_1=5$, the energy differences were always below $5\times 10^{-3}$, indicating that the $N_1=4$ calculations are already well converged. 

\section{Ground-state energy dependence on Ising SOC}
\label{sec:App_energy_vs_lambda}
\begin{figure}[h!]
  \centering
  \includegraphics[width=0.4\columnwidth]{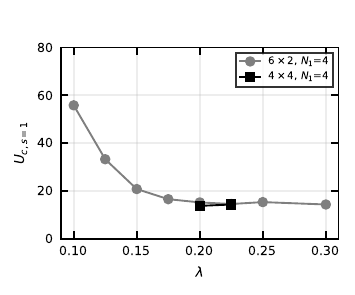}
  \caption{Estimate of the critical interaction strength $U_c$ for the ferromagnetic transition in terms of the spin gap at $\nu=1$, as a function of $\lambda$. We performed most of the calculations for the 12-unit-cell lattice, but also show results for the 16-unit-cell for some values of $\lambda$ to show that the results are well converged. }
  \label{fig:App_convergence_lambda}
\end{figure}

In this Appendix section, we inspect the ferromagnetic transition and the ground-state energy dependence on $S_z$ for different values of the Ising SOC $\lambda$.
To inspect the ferromagnetic transition, we evaluate the critical interaction strength $U_c$ above which the spin gap becomes positive. Although this is only a lower-bound for the ferromagnetic transition, it clearly shows its trend as a function of $\lambda$, and is a calculation that can be well converged in system size. The results are shown in Fig.~\ref{fig:App_convergence_lambda}, where it can be seen that $U_c$ decreases with $\lambda$ from $\lambda=0.1$ to $\lambda=0.2$, approximately stabilizing at $\lambda \gtrsim 0.2$.

\begin{figure}[h!]
  \centering
  \includegraphics[width=\columnwidth]{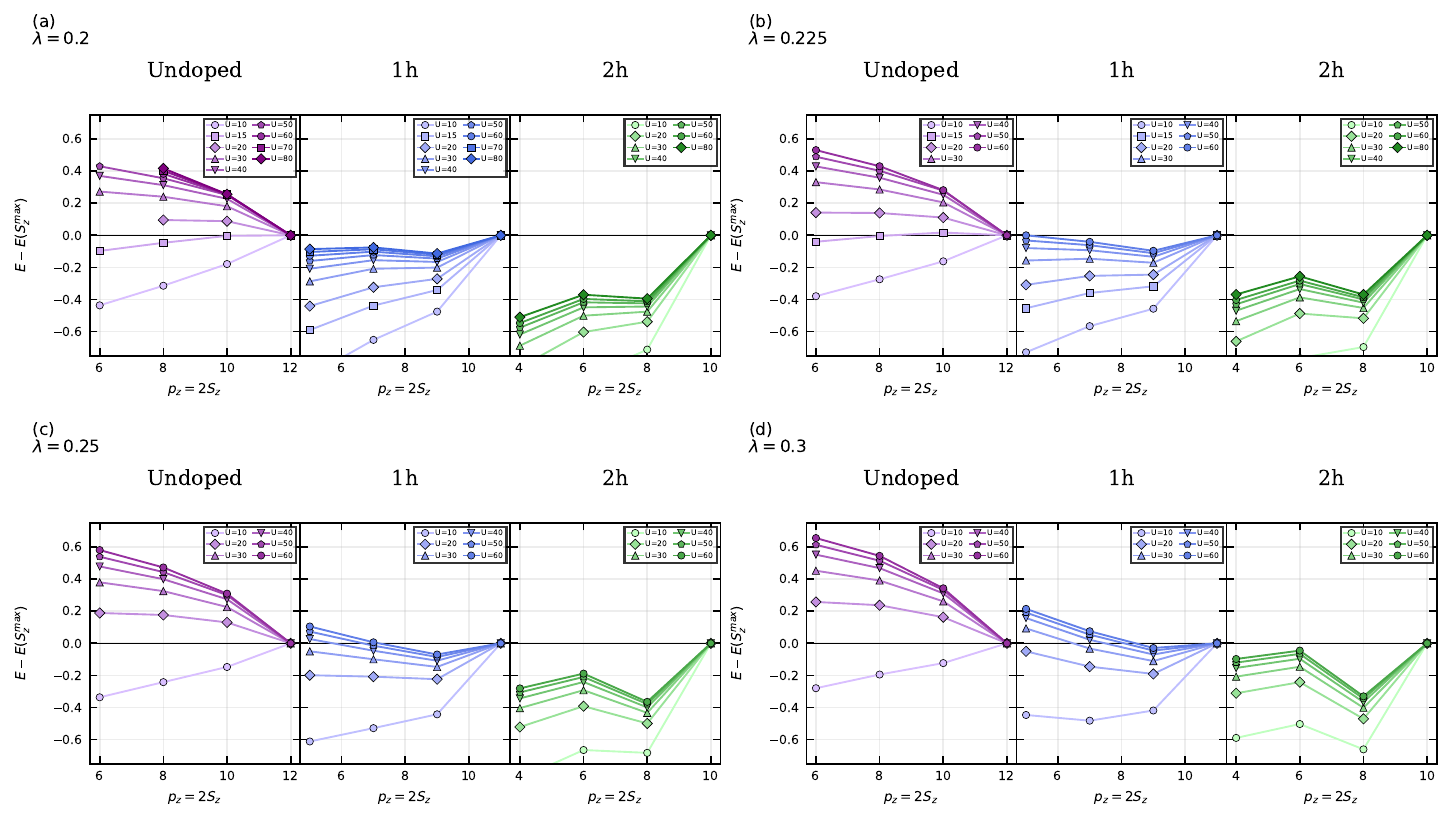}
  \caption{Energy difference with respect to the fully polarized state as a function of $p_z=2S_z$ for different interaction strength $U$, for a 16-unit-cell lattice and $N_1=4$. Different panels correspond to different values of $\lambda$, indicated close to each label. }
  \label{fig:App_varLambda}
\end{figure}

We  also inspect the ground-state energy dependence on $S_z$ for different values of the Ising SOC $\lambda$. The results are shown in Fig.$\,$\ref{fig:App_varLambda} where we show that increasing $\lambda$ stabilizes the \textit{2h1s} skyrmion-bipolaron as the ground-state at a large enough $U$. For smaller $\lambda$ and $U$, the ground-state occurs in a smaller spin sector, suggesting that it is energetically favorable for the two holes to bind with a higher number of spin-flips. 

\section{Binding energy calculations}

We now turn to explicit calculations of binding energies for the skyrmion-polaron and skyrmion-bipolaron bound-states. We start by defining the energy of an excitation of $n$ holes and $m$ spin flips with respect to the fully polarized Chern ferromagnet as 

\begin{equation}
  \epsilon(\textrm{n}h\textrm{m}s) = E(N-\textrm{n},(N-\textrm{n})/2-\textrm{m}) - E(N,N/2)\,.
\end{equation}

\noindent where once again $E(N,S_z)$ is the ground-state energy for $N$ particles and spin sector $S_z$.   Using this, we define the binding energy of the skyrmion-polaron bound-state as 

\begin{equation}
  \Delta_B(1h1s) = \epsilon(1h1s) - \epsilon(1h) - \epsilon(1s)\, ,
\end{equation}
and the binding energies for the skyrmion-bipolaron bound-state as
\begin{equation}
  \begin{aligned}
    \Delta_B^{(1)}(2h1s) &= \epsilon(2h1s) - \epsilon(1h1s) - \epsilon(1h)\,, \\
    \Delta_B^{(2)}(2h1s) &= \epsilon(2h1s) - \epsilon(2h) - \epsilon(1s)\,.
  \end{aligned}
\end{equation}

\begin{figure}[h!]
  \centering
  \includegraphics[width=\columnwidth]{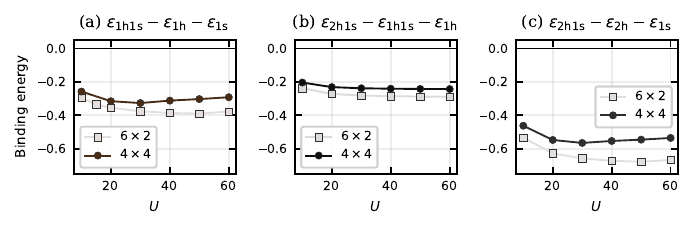}
  \caption{ Binding energies for the skyrmion-polaron (a) and skyrmion-bipolaron (b,c) bound states as a function of interaction strength $U$ for $\lambda=0.225$, $N_1=4$ and different system sizes indicated in the figures. }
  \label{fig:binding_energy}
\end{figure}

We plot all these binding energies as a function of $U$ in Fig.~\ref{fig:binding_energy}. We can see that even though there is still some finite-size dependence, the binding energies are always negative for all studied interaction strengths. This shows strong evidence for the formation of the skyrmion-polaron and skyrmion-bipolaron bound states.

\section{Momentum-resolved energy spectra}
\begin{figure}[h!]
  \centering
  \includegraphics[width=0.8\columnwidth]{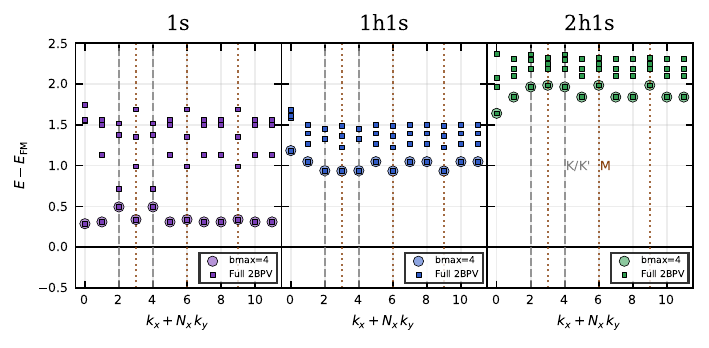}
  \caption{Momentum-resolved energy spectrum of the four lowest-energy modes for $\lambda=0.225$ and $U=100$, for the magnon (a), skyrmion-polaron (b) and skyrmion-bipolaron (c). The results are for a 12-unit-cell lattice. The full two-band results are shown together with the results for $N_1=4$ (for which we only show the lowest-energy mode per momentum), to show the good convergence of the latter. }
  \label{fig:App_dispersion}
\end{figure}

In the main text, we showed the energy dispersion for the skyrmion-bipolaron bound state. In this Appendix section we complement these results by showing in Fig.$\,$\ref{fig:App_dispersion} the momentum-resolved energy spectrum of the lowest-energy modes for the magnon, skyrmion-polaron and skyrmion-bipolaron excitations. 

\section{Additional DMRG results}
In this section we provide additional DMRG results to support the discussions in the main text.
\subsection{Spin chirality flip by FM background}
\begin{figure}[h!]
  \centering
  \includegraphics[width=\columnwidth]{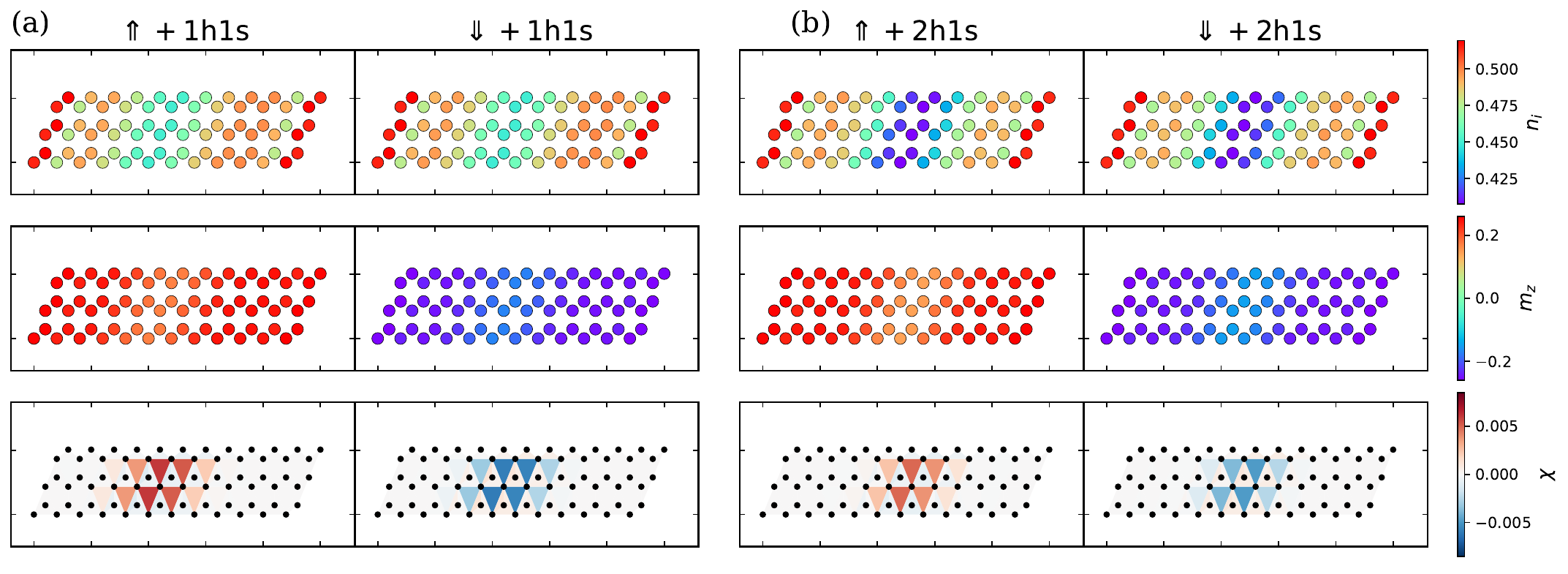}
  \caption{DMRG results for the charge density, magnetization and spin chirality for $\lambda=0.225$ and $U=100$. Ground-state upon flipping one spin and doping one hole (a) or two holes (b) with respect to the $\ket{\Uparrow}$ and $\ket{\Downarrow}$ Chern ferromagnet states. }
  \label{fig:App_spinChiralityFlip}
\end{figure}

We start by showing explicitly in Fig.~\ref{fig:App_spinChiralityFlip} that changing the polarization of the parent Chern ferromagnet changes the spin chirality of the skyrmion-polarons and skyrmion-bipolarons. For the skyrmion-polaron in Fig.~\ref{fig:App_spinChiralityFlip}(a), it can be seen that the region where the hole is localized (where there is a depletion of charge density) is the same where the magnetization changes with respect to the background value and where a finite spin-chirality emerges. The sign of the spin-chirality is set by the polarization/Chern number of the parent Chern ferromagnet. Similar results are shown for the skyrmion-bipolaron in Fig.~\ref{fig:App_spinChiralityFlip}(b).

\subsection{Skyrmion-bipolaron interaction vs Ising SOC}
\begin{figure}[h!]
  \centering
  \includegraphics[width=\columnwidth]{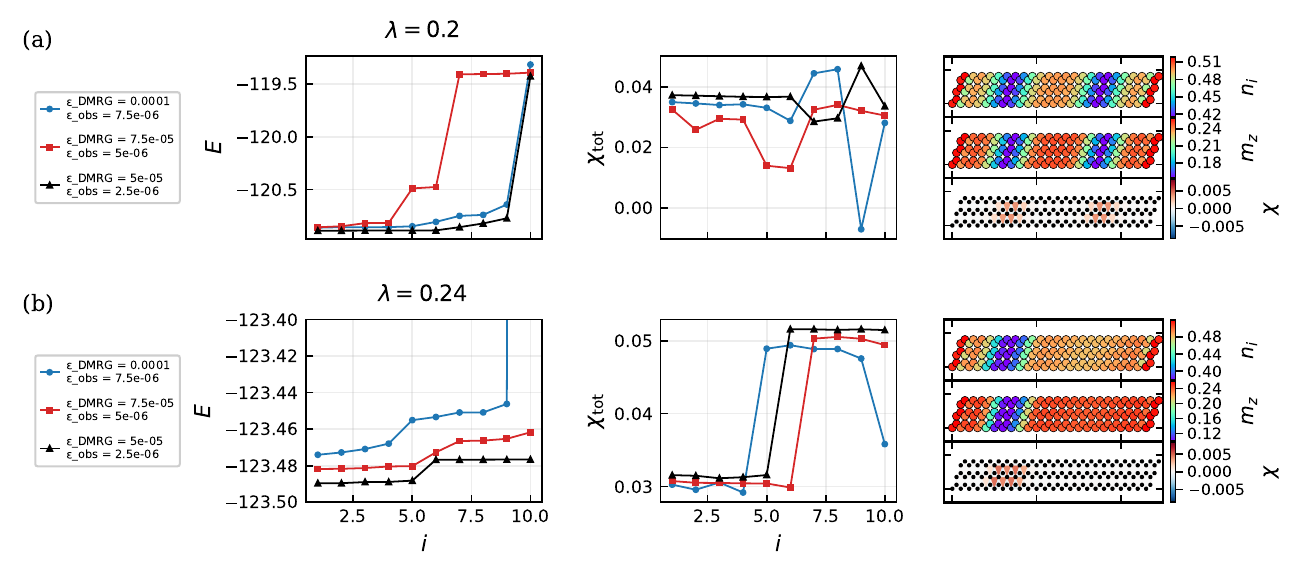}
  \caption{DMRG results for two spin-flips and four doped holes with respect to the fully polarized Chern ferromagnet, for $U=100$ and (a) $\lambda=0.2$; (b) $\lambda=0.24$, respectively below and above the binding transition for the skyrmion-bipolarons. The leftmost panels correspond to the ground-state energies obtained for all DMRG runs using 10 different random initial states and different convergence criteria indicated in the figure. The corresponding total spin chiralities are shown in the middle panels. The rightmost panels show the real-space charge density, magnetization and spin chirality for the lowest-energy state under the finer convergence criterion. }
  \label{fig:App_skyrmionBipolaronInteraction}
\end{figure}

In this Appendix section we explore in more detail the unbinding-binding transition for the skyrmion-bipolarons for different values of Ising SOC $\lambda$.
In Fig.~\ref{fig:App_skyrmionBipolaronInteraction} we show results for $\lambda<\lambda_{c,2}$ and $\lambda>\lambda_{c,2}$, respectively below and above the unbinding transition. The DMRG results for the 10 different random initial states show that there is a plateau of low-energy states with quasi-degenerate energies that is stable to improving the convergence criterion. This indicates that the algorithm is converging to the right ground-state and not to a local minimum -- we checked that all the quasi-degenerate lowest-energy states had the same density, magnetization and spin-chirality profiles. It is interesting to note that for the $\lambda>\lambda_{c,2}$ example, we can see that some of the DMRG runs converged to states with slightly higher energy with a larger total chirality. We verified that these correspond to the states with two unbound skyrmion-bipolarons (not shown).

\end{document}